\documentclass[english]{procmult}
\usepackage{graphicx, babel, mathrsfs, bbm, amsmath, amssymb, bm}

\begin{document}

\title{Fluid Entropy as Time-Variable in Canonical Quantum Gravity}
\author{Francesco Cianfrani$^\dag$, Giovanni Montani$^{\dag\ddag}$ \and Simone Zonetti$^\S$}
\institute{
$^\dag$  {Dipartimento di Fisica, Universit\`a degli Studi di Roma ``Sapienza'',\\
Piazzale Aldo Moro 5, 00185 Rome, Italy.}\\
$^\ddag$  {ENEA C.R. Frascati (Dipartimento F.P.N.), via Enrico Fermi 45,\\ 
00044 Frascati, Rome, Italy, \\ ICRANET C. C. Pescara, Piazzale della Repubblica, 10, 65100 Pescara, Italy.}\\
$^\S$   {Center for Particle Physics and Phenomenology (CP3)\\
Université catholique de Louvain\\
Chemin du Cyclotron, 2
B-1348 Louvain-la-Neuve,
Belgium .}}

\maketitle
\abstract{The Brown-Kucha\v r mechanism is applied in the case of General Relativity coupled with the Schutz' model for a perfect fluid. Using the canonical formalism and manipulating the set of modified constraints one is able to recover the definition of a time evolution operator, i.e. a physical Hamiltonian, expressed as a functional of gravitational variables and the entropy.
}

\section{Introduction}

The problem of time \cite{Isham:1992a} is a well known issue in the context of the canonical approaches to Quantum Gravity \cite{rovelli:2004} \cite{rovelli:1997}, and is a straightforward consequence of the vanishing Hamiltonian that characterizes diff-invariant systems. 
Among the different approaches have been developed, trying to recover a meaningful definition of time \cite{Isham:1992a}  \cite{ashtekar:1987} \cite{montani:2007} \cite{cianfrani:2008} \cite{cianfrani:2007} a possible solution is to couple matter to the gravitational field. Using the strong duality between matter and reference frames (\cite{thiemann:2006} \cite{cianfrani:2008} \cite{mercuri:2003}, see \cite{kuchartorre:1991.1} and \cite{kuchartorre:1991.2} for a clear example) one can intuitively use fluid particles to label space-time points, while general covariance can be recovered simply re-parameterizing the coordinate dependence. This duality can be used in a more convenient way through the Brown-Kucha\v r mechanism \cite{kucharbrown:1995}, which, by manipulating the set of constraints of General Relativity coupled with matter, enables one to recover the definition of a meaningful, non vanishing, evolution operator.\\
Here this procedure is applied to the Schutz' model of a perfect fluid \cite{schutz:1971}, 
which provides a proper representation of a fluid constituted by particles whose own entropy is conserved. 


In particular, it is outlined that the Brown-Kucha\v r mechanism works also in this case and it gives a strong temporal value to the entropy field associated with this particular matter field. This is an intriguing result in view of finding a connection between the thermodynamical time and time in Quantum Gravity.\\

This work consists in five more sections.
: in section 2 the Brown-Kucha\v r mechanism will be reviewed in the case of a generic scalar field Lagrangian, while in section 3 the Schutz' fluid will be described in the thermodynamical context. In section 4 the Hamiltonian formulation for the uncoupled model will be presented, and in section 5 the mechanism will be finally applied to the coupled model. In section 6 results and perspectives are spotted out.\\

\section{The Brown-Kucha\v r Mechanism}\label{thekbmechanism}
Here the Brown-Kucha\v r Mechanism is reviewed in the case of a real scalar field, with Lagrangian function\footnote{The metric is assumed to have signature $(-,+,+,+)$} in the form $\mathcal{L}_F = \mathcal{L}_F [-\phi_{,\mu}\phi^{,\mu}] = \mathcal{L}_F [\Upsilon]$ \footnote{the comma denotes the ordinary derivative and space-time indices (greek letters) are contracted with the space-time metric.}. To keep the general covariance of the theory untouched one performs the standard 3+1 ADM splitting of the space-time \cite{ADM:1962} \cite{thiemann:2001}, and computes the conjugate momentum tp $\phi$:
\begin{equation}
\pi = \frac{\delta \mathcal{S}_F}{\delta \dot{\phi}} = -2\sqrt{q} \phi_n \frac{\delta \mathcal{L}_F[\Upsilon]}{\delta \Upsilon},
\end{equation}
where one has defined $q = det(q_{ab})$. This definition, together with the splitting for $\Upsilon$, allows to complete the Legendre transformation, so that the Hamiltonian will contain only spatial quantities and the conjugate momentum $\pi$, and will take the simple form $\mathcal{H}_F = \int d^3 \! x (H^F N + H^F_a N^a)$.\\
By adding the Einstein-Hilbert action of GR the Hamiltonian density will simply be the sum of the matter-free super-Hamiltonian and super-momentum with the scalar field terms defined above, because there are no derivatives of the metric tensor $g_{\mu \nu}$ in the field action.
The modified constraints of GR will read $H = H^G[q,P] + H^F[\pi, V, q]$ and $H_a = H^G_a [q,P] + \phi_a \pi$.\\
At this point one can manipulate the set of constraints in order to obtain an equivalent Super-Hamiltonian, in the form:
\begin{equation}\label{eqconstraint}
\pi - h[ q, P] = 0,
\end{equation}
which can be seen as a Schr\"odinger equation for observables\footnote{Here one assumes that observables exist, and have the standard property of vanishing Poisson brackets with the whole set of constraints.}, as soon as the field $\phi$ is takes as an internal time for the system. The physical Hamiltonian will simply consist in $\mathcal{H}_{phys} = \int d^3 \! x h(x)$, under the necessary conditions: (a) Independence from $\phi$, (b) Invariance under the 3-diff group. This is manifest if $h$ is a scalar density of weight one, (c) Invariance under re-parametrization of the time-like variable, {\it i.e.} $\{h,h\}=0$ strongly \cite{kucharbrown:1995}.

\section{Schutz' Perfect Fluid}

The equation of state for the Schutz' perfect fluid can be obtained directly from thermodynamical principles, when dealing with a baryonic perfect fluid (\cite{schutz:1970} \cite{schutz:1971}  \cite{fermi:1936}) and reads $p=\rho_0 ( \mu - TS )$, where the specific inertial mass $\mu$ can be fixed with the normalization condition for the 4-velocity, i.e. $\mu = \sqrt{-v^\mu v_\mu}$. Moreover the normalized 4-velocity can be written as a combination of scalar fields and their gradients:
\begin{equation}\label{4velocity}
U_\nu = \mu^{-1}(\phi_{,\nu} + \alpha \beta_{,\nu} + \theta S_{,\nu}) = \frac{v_\nu}{\mu}.
\end{equation}
Here $S$ is the entropy per baryon, while the other fields have no direct physical meaning. From now on the commas denoting fluid gradients will be dropped with no ambiguities.\\
Using the stress-energy tensor and the Einstein equations for matter $\delta \mathcal{L}_F /\delta g_{\mu \nu} = T^{\mu \nu}/2$, one can compute the Lagrangian density for the fluid. With such an integration one can identify $\mathcal{L}_F=\sqrt{-g}p = \sqrt{-g}\rho_0 ( \sqrt{-v^\mu v_\mu}-TS )$.
\subsection{Hamiltonian formulation}
The fluid action, after the ADM splitting, takes the form\footnote{Latin indices denotes spatial quantities, and are contracted with the spatial metric $q^{ab}$}:
\begin{equation}\label{schutzADMaction}
\mathcal{S}_F=\int_\mathbbm{R} \! dt \int_\sigma \! d^3 \ x  \sqrt{q} N \rho_0 \left( \sqrt{(v_\mu n^\mu)^2 - v_a v^a} - TS \right).
\end{equation}

From the definition of momenta the following constraints are inferred 
\begin{equation}
\chi_1=p_{\alpha} = 0, \qquad
\chi_2=p_{\beta}-  \alpha \pi=0, \qquad
\chi_3=p_{\theta} = 0, \qquad
\chi_4=p_{S}- \theta \pi.\label{sconstraint}
\end{equation}

Calculating the Poisson brackets matrix one can easily see that such constraints are second class, and the rank of the matrix is maximum.

The Legendre transformation will lead to the formal Hamiltonian function:
\begin{equation}
H=N\left(\sqrt{(\pi^2-q\rho_0^2)V}+q\rho_0TS\right)+N^a\pi v_a,
\end{equation}
where $V = v_a v^a$ and non-spatial quantities have been eliminated using the definition of $\pi$ and the splitting of $v^\mu v_\mu$.

\section{Coupling with General Relativity}
By coupling the fluid with GR one expects the primary constraints to be simply the union of the two sets of primary constraints, since the fluid action does not contain any derivative of the metric tensor $g$. The secondary constraints will be identified with the functionals that appear enforced by the lapse function and the shift vector, which are:
\begin{eqnarray}\label{sgrmultipleH}
H = \sqrt{V(\pi^2 - q \rho_0^2)}+ \sqrt{q} \rho_0 S T + H^G,\qquad
H_a = \pi v_a + H^G_a.
\end{eqnarray}
The algebra of these constraints is not modified by the presence of the Schultz fluid. 
Therefore \emph{the full covariance of model is preserved} and the only secondary constraints that appear are the super-Hamiltonian and super-momentum. So it can be shown that \emph{they preserve their role of generators of the diffeomorphisms and exhibit a closed algebra}.\\
This means that there are no tertiary constraints in the model, because the consistency conditions are fulfilled. The resulting Hamiltonian is vanishing as a consequence of the fact that the diffeomorphism invariance was never broken.
At this point one is able to apply the Brown-Kucha\v r mechanism: squaring the super-momentum and imposing it on the super-Hamiltonian, one can solve the latter for $\pi$, obtaining something in the form $\pi - h = 0$. 
This way one gets
\begin{equation}\label{sgrhfunction}
\pi\pm\sqrt{\rho_0q\frac{d}{d-\Xi^2}}=0 = \pi  - h,
\end{equation}
where $\Xi = \sqrt{q} \rho_0 S T + H^G$ and $d = H^G_a H^G_b q^{ab}$. $\Xi$ is also the expression the super-Hamiltonian reduces to in the co-moving frame (i.e. $N=1$ and $N_a=0$).\\
The $h$ function is now the candidate for the construction of the physical Hamiltonian: it is a scalar density of weight one, 
$h$ does not contain the field $\phi$, so its Poisson brackets with $\pi$ are vanishing, and the commutator $\{h(f), h(g)\}$ does not contain any dependence on $\pi$. 
Hence, $\{h(f), h(g)\}$ cannot reproduce any other constraint. Since the full system of constraints must 
be first-class ($H$ and $\pi - h$ are equivalent constraints) one can conclude, according with 
\cite{kucharbrown:1995}, that $\{h(f), h(g)\} \simeq 0$ (strongly vanishes), for arbitrary smearing fields $f(x)$ and $g(x)$.
Finally, since the $h$ and $H_a$ generate respectively re-parametrization of the time-like variable and spatial diffeomorphisms, $h$ itself is an observable.

The direct interpretation of entropy as time variable comes out in the particular case of the co-moving frame (i.e. $N=1$ and $N_a=0$), where the conjugate momentum reduces to $\pi = - \sqrt{q} \rho_0$, while the set of secondary constraints become 
$\Xi = \sqrt{q} \rho_0 S T + H^G=0$ and $\Xi_a = H^G_a=0$.\\
One can use the super-Hamiltonian to obtain an expression for $\sqrt{q} \rho_0$, and recalling the expression for the momentum conjugate to $S$ \eqref{sconstraint} one can multiply both sides of \eqref{sgrhfunction} by $\theta$, so that:
\begin{equation}
S p_S = \frac{\theta H^G}{T} = \bar{h},
\end{equation}
which integrated over the spatial manifold allows to \emph{identify the time parameter $\tau$ with the logarithm of the entropy per baryon}, and finally see such equation as a Schr\"odinger-like evolution equation for observables. 

\section{Concluding remarks}
A possible solution to the problem of time in canonical quantum gravity is proposed: using the Brown-Kucha\v r mechanism one is able to write an equivalent set of secondary constraints for GR coupled with the Schutz' perfect fluid. In this context the super-Hamiltonian acquires the form of a Schr\" odinger operator, i.e. becomes a physical Hamiltonian, which generates time evolution for observables. Its expression is complex, but the entropy field $S$ enters directly its definition: time and entropy are highly related.\\
This relation is much more evident in the co-moving frame, where the absence of the spatial gradients of the fluid allows one to find a clearer physical interpretation: in this particular case the logarithm of entropy per baryon is exactly the time variable.
Considering that the number of particles $n$ for the fluid is fixed \cite{schutz:1970}, one can simply obtain the entropy of the system by multiplying by $n$ the $S$ field. This is surprising, because there are no hints in the theory about that privileged role played by the entropy, and because $S$, and so its logarithm, are naturally future pointing quantities in a closed system.\\
Moreover the evolution operator is rather simple, above all because the matter-free super-Hamiltonian appears linearly in the physical Hamiltonian. The only fluid variable is $\theta$, whose equation of motion contains the arbitrary temperature function $T$, and whose conjugate momentum is vanishing. It is just an arbitrary field fixed by initial value conditions.

\section*{Aknowledgements}
We would like to thank M. Testa for his precious suggestions about the preservation of the diff-invariance in the coupled theory. The work of SZ is supported by the Belgian Federal Office for Scientific, Technical and Cultural Affairs through the Interuniversity Attraction Pole P6/11.

\end{document}